\documentclass[10pt,a4paper,twoside,dvipdfm]{aa}
\usepackage{epsfig}
\usepackage{baltlat6}
\usepackage{array}
\usepackage{here}
\usepackage{natbib}
\usepackage[dvipdfm=true]{hyperref}
\hypersetup{pdftitle = The title of my PDF, pdfauthor = My name,
pdfsubject= The subject, pdfkeywords = keyword1 keyword2 keyword3}
\hypersetup{colorlinks = true, linkcolor = blue, anchorcolor = red,
citecolor = blue, filecolor = red, pagecolor = red, urlcolor = blue}

\begin{document}
\ \
\vspace{0.5mm}
\setcounter{page}{37}
\vspace{8mm}

\titlehead{Baltic Astronomy, vol.\,23, 37--54, 2014}

\titleb{HOW DOES THE STRUCTURE OF SPHERICAL DARK MATTER HALOES AFFECT
THE TYPES OF ORBITS IN DISK GALAXIES?}

\begin{authorl}
\authorb{Euaggelos E. Zotos}{}
\end{authorl}

\moveright-3.2mm
\vbox{
\begin{addressl}
\addressb{}{Department  of Physics, School of Science, Aristotle
University of Thessaloniki, \\ GR-541 24, Thessaloniki, Greece;
e-mail: evzotos@physics.auth.gr}
\end{addressl} }

\submitb{Received: 2014 March 24; accepted: 2014 March 30}

\begin{summary} The main objective of this work is to determine the
character of orbits of stars moving in the meridional $(R,z)$ plane of
an axially symmetric time-independent disk galaxy model with a central
massive nucleus and an additional spherical dark matter halo component.
In particular, we try to reveal the influence of the scale length of the
dark matter halo on the different families of orbits of stars, by
monitoring how the percentage of chaotic orbits, as well as the
percentages of orbits of the main regular resonant families evolve when
this parameter varies.  The smaller alignment index (SALI) was computed
by numerically integrating the equations of motion as well as the
variational equations to extensive samples of orbits in order to
distinguish safely between ordered and chaotic motion.  In addition, a
method based on the concept of spectral dynamics that utilizes the
Fourier transform of the time series of each coordinate is used to
identify the various families of regular orbits and also to recognize
the secondary resonances that bifurcate from them.  Our numerical
computations reveal that when the dark matter halo is highly
concentrated, that is when the scale length has low values the vast
majority of star orbits move in regular orbits, while on the oth er hand
in less concentrated dark matter haloes the percentage of chaos
increases significantly.  We also compared our results with early
related work.  \end{summary}

\begin{keywords}
galaxies: kinematics and dynamics, structure; chaos; dark matter haloes
\end{keywords}

\resthead{Spherical dark matter haloes and star orbits}
{Euaggelos E. Zotos}

\sectionb{1}{INTRODUCTION}

A large amount of observational data indicate that disk galaxies are
often surrounded by massive and extended dark matter haloes.
Undoubtedly the best tool to study dark matter haloes in galaxies is the
galactic rotation curve derived from neutral hydrogen HI (e.g.,
\citealt{C85,PS95,HS97}).  The determination of the exact shape of a
dark matter halo however, is a challenging task.  Numerical simulations
suggest that dark matter haloes are not only spherical, but may also be
prolate, oblate, or even triaxial (e.g.,
\citealt{MF96,C00,KTT00,OM00,JS02,WBPKD02,KE05,AFP06,CLMV07,WMJ09,EB09}).
The variety of the shapes of galactic haloes points out that the
structure of these objects plays an important role in the orbital
behavior and, generally, in the dynamics of a galaxy.

The presence of dark matter haloes in galaxies is indeed expected by the
standard cold dark matter (CDM) cosmology models regarding formation of
galaxies.  The most well-known model for CDM haloes is the flattened
cuspy Navarro Frenk White (NFW) model (\citealt{NFW96,NFW97}), which is
simplified to be spherical.  Most CDM models, however, take considerable
deviations from the standard spherically symmetric dark matter halo
distributions into account.  For instance, the model of formation of
dark matter haloes in a universe dominated by CDM by \citet{FWDE88}
produced triaxial haloes with a preference for prolate configurations.
In addition, numerical simulations of dark matter halo formation
conducted by \citet{DC91} are consistent with haloes that are triaxial
and flat.  There are roughly equal numbers of dark haloes with oblate
and prolate forms.

Over the last years several dynamical models have been developed in
order to model the properties of dark matter haloes.  A
three-dimensional model consisting of a disk, a spherical nucleus, and a
logarithmic asymmetric dark matter halo component was used in
\citet{CZ09}.  For simplicity, a nearly spherical dark matter halo with
an internal, small deviation from spherical symmetry described by the
term $-\lambda x^3$ was chosen.  The results of this work suggest that
even small asymmetries in the galactic halo play a significant role in
the nature of three-dimensional orbits, mainly by depopulating the box
family; the box orbits become chaotic as the value of the internal
perturbation increases.  In the same vein, a similar three-dimensional
composite galaxy model was utilized in \citet{CZ11}, however, in this
case the dark mater halo was modeled by a mass Plummer potential.  The
mass of the halo was found to be an important physical quantity, acting
as a chaos controller in galaxies.  In particular, the percentage of
chaotic orbits reduces rapidly as the mass of the spherical dark halo
increases.  Moreover, it was found that the amount of chaos is higher in
asymmetric triaxial galaxies when they are surrounded by less
concentrated spherical dark halo components.

Furthermore, in two earlier papers (\citealt{C97,PC06})
two-dimensional, axially symmetric or non-axially symmetric active
galaxy models with an additional spherical halo component were studied.
In both cases it was observed that the presence of a spherical halo
resulted in reduced area in the phase space occupied by the chaotic
orbits.  Moreover, the behavior of orbits in an active galaxy with a
biaxial (prolate or oblate) dark matter halo was investigated recently
in \citet{CZ10}.  In this work, we studied how the regular or chaotic
nature of orbits is influenced by some important quantities of the
system, such as the flattening parameter of the halo, the scale length
of the halo component, and the conserved component of the angular
momentum.  It was found that when a biaxial halo component is present
there is a linear relationship between the chaotic percentage in the
phase plane and the flattening parameter.  In contrast, the relation
between chaos and the scale length of the halo was found to be not
linear  but exponential.  A similar model was used in \citet{H04} for
numerical simulations of the evolution of a system like the Sagittarius
dSph in a variety of galactic potentials varying the flattening
parameter.

In a very recent paper \citet{Z14} we revealed how all the dynamical
parameters of a multi component model influence the percentages of
orbits.  For the description of the properties of the dark matter halo
we used a softened logarithmic potential.  Usually , logarithmic
potentials are utilized for modeling dark matter haloes, however, in the
present work we decided to use a mass type potential.  We made this
choice for two main reasons:  (i) we wanted to know the exact amount of
mass of the halo; logarithmic potentials cannot give this information
and (ii) it would be of particular interest to compare the corresponding
results derived using two different types of model potentials for the
dark matter halo.

The layout of the paper is as follows:  Section 2 contains a detailed
presentation of the structure and the properties of our galactic
gravitational model.  In Section 3 we describe the computational
methods we used in order to determine the character and the
classification of orbits.  In the following section, we investigate how
the scale length of the dark matter halo influences the nature as well
as the evolution of the percentages of the different families of orbits.
The paper ends with Section 5, where the main conclusions of our
numerical analysis and the discussion are presented.

\sectionb{2}{PROPERTIES OF THE GALACTIC MODEL}

In this investigation we try to reveal the regular or chaotic nature of
orbits of stars moving in the meridional plane of an axially symmetric
disk galaxy with a central massive nucleus and a spherical dark matter
halo.  We use the usual cylindrical coordinates $(R, \phi, z)$, where
$z$ is the axis of symmetry.

The total gravitational potential $\Phi(R,z)$ in our model consists of
three components:  the central nuclear component $\Phi_{\rm n}$, the
disk potential $\Phi_{\rm d}$ and the spherical dark matter halo
component $\Phi_{\rm h}$.  For the description of the central nucleus,
we use a Plummer potential (e.g., \citealt{BT08})
\begin{equation}
\Phi_{\rm n}(R,z) = \frac{- G M_{\rm n}}{\sqrt{R^2 + z^2 + c_{\rm n}^2}}.
\label{Vn}
\end{equation}
Here $G$ is the gravitational constant, while $M_{\rm n}$ and $c_{\rm
n}$ are the mass and the scale length of the nucleus, respectively.
This potential has been used successfully in the past to model and,
therefore, interpret the effects of the central mass component in a
galaxy (see e.g., \citealt{HN90,HPN93,Z11b,Z12,ZC13,Z14}).  At this
point, we must emphasize that we do not include any relativistic
effects, because the nucleus represents a bulge rather than a black hole
or any other compact object.

On the other hand, the galactic disk is represented by the well-known
Miyamoto-Nagai potential \citet{MN75}
\begin{equation}
\Phi_{\rm d}(R,z) = \frac{- G M_{\rm d}}{\sqrt{R^2 + \left(\alpha + \sqrt{h^2 + z^2}\right)^2}},
\end{equation}
where, $M_{\rm d}$ is the mass of the  disk, $\alpha$ is the scale
length of the disk, and $h$ corresponds  to the disk's scale height
(e.g., \citealt{Z11a}). Furthermore, the   dark matter halo is modeled
by another Plummer potential
\begin{equation}
\Phi_{\rm h}(R,z) = \frac{- G M_{\rm h}}{\sqrt{R^2 + z^2 + c_{\rm h}^2}},
\label{Vh}
\end{equation}
where $M_{\rm h}$ and $c_{\rm h}$ are the  mass and the scale length of
the dark halo, respectively.

We use a system of galactic units where the unit of length is 1 kpc, the
unit of velocity is 10 km s$^{-1}$, and $G = 1$.  Thus, the unit of mass
is $2.325 \times 10^7 {\rm M}_\odot$, that of time is $0.9778 \times
10^8$ yr, the unit of angular momentum (per unit mass) is 10 km
kpc$^{-1}$ s$^{-1}$, and the unit of energy (per unit mass) is 100
km$^2$s$^{-2}$.  In these units, the values of the involved parameters
are:  $M_{\rm n} = 400$ (corresponding to $9.3\times 10^{9}$ M$_\odot$),
$c_{\rm n} = 0.25$, $M_{\rm d} = 7000$ (corresponding to $1.63\times
10^{11}$ M$_\odot)$, $\alpha = 3$, $h = 0.175$ and $M_{\rm h} = 10000$
(corresponding to $2.325\times 10^{11}$ M$_\odot$).  The particular
values of the parameters were chosen with a Milky Way-type galaxy in
mind (e.g., \citealt{AS91}).  The above-mentioned set of values of the
parameters, which are kept constant throughout the numerical
calculations, secures positive mass density everywhere and is free of
singularities.  The scale length of the dark matter halo, on the other
hand, is treated as a parameter.


\begin{figure}[!t]
\centerline{\includegraphics[width=0.7\hsize]{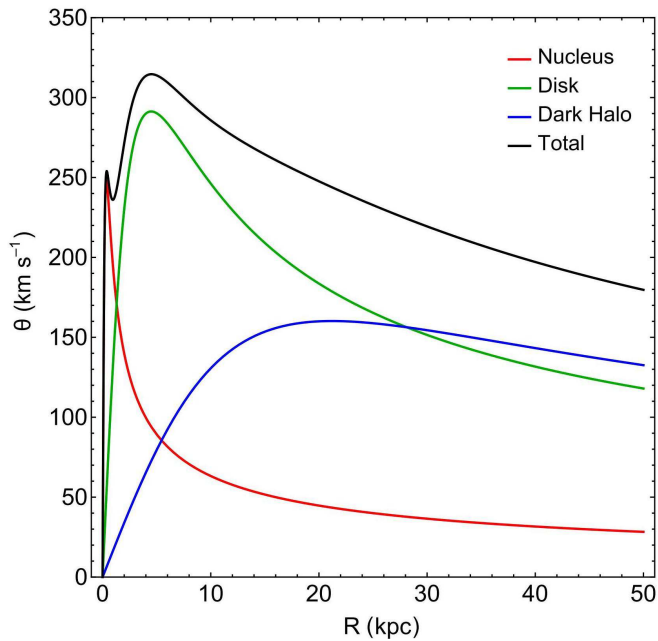}}
 \caption{A plot of the rotation curve in our galactic model. We
 distinguish the total circular velocity (black) and also the
 contributions from the central massive nucleus (red), the disk (green)
and that of the dark matter halo (blue).}
\label{rotvel}
\end{figure}

Undoubtedly, one  of the most important physical quantities in galaxies
is the circular  velocity in the galactic plane $(z = 0)$
\citep[e.g.,][]{Z11c} which is defined as,
\begin{equation}
\theta(R) = \sqrt{R\left|\frac{\partial \Phi(R,z)}{\partial R}\right|_{z = 0}}.
\label{rcur}
\end{equation}
In Fig.~\ref{rotvel} we present a plot of $\theta(R)$ (black curve) for
our galactic model when $c_{\rm h} = 15$.  Moreover, in the same plot,
the red line shows the contribution from the central nucleus, the green
curve is the contribution from the disk, while the blue line corresponds
to the contribution form the dark matter halo.  It is seen that each
contribution prevails in different distances form the galactic center.
In particular, at small distances, when $R\leq 2$ kpc, the contribution
from the massive central nucleus dominates, while at mediocre
distances, $2 < R < 28$ kpc, the disk contribution is the dominant
factor.  On the other hand, at very large galactocentric distances, $R >
28$ kpc, we see that the contribution from the dark matter halo
prevails, thus forcing the rotation curve to remain flat with increasing
distance from the center.  We also observe the characteristic local
minimum (at $R \simeq 2$ kpc) of the rotation curve due to the massive
nucleus, which appears when fitting the observed data to a galactic
model (e.g., \citealt{GHBL10,IWTS13}).

Exploiting the fact that the $L_z$-component of the total angular
momentum $L_z$ is conserved because the gravitational potential
$\Phi(R,z)$ is axisymmetric, orbits can be described by means of the
effective potential
\begin{equation}
\Phi_{\rm eff}(R,z) = \Phi(R,z) + \frac{L_z^2}{2R^2}.
\label{veff}
\end{equation}

Then, the basic equations of motion on the meridional plane are
\begin{equation}
\ddot{R} = - \frac{\partial \Phi_{\rm eff}}{\partial R}, \ \ \ \ddot{z} = - \frac{\partial \Phi_{\rm eff}}{\partial z},
\label{eqmot}
\end{equation}
while the equations governing  the evolution of a deviation vector
${\bf{w}} = (\delta R, \delta z,  \delta \dot{R}, \delta \dot{z})$,
which joins the corresponding phase space  points of two initially
nearby orbits, needed for the
calculation of the standard indicators   of chaos (the SALI in our
case), are given by the variational equations
\begin{eqnarray}
\dot{(\delta R)} &=& \delta \dot{R}, \ \ \ \dot{(\delta z)} = \delta \dot{z}, \nonumber \\
(\dot{\delta \dot{R}}) &=&
- \frac{\partial^2 \Phi_{\rm eff}}{\partial R^2} \delta R
- \frac{\partial^2 \Phi_{\rm eff}}{\partial R \partial z}\delta z,
\nonumber \\
(\dot{\delta \dot{z}}) &=&
- \frac{\partial^2 \Phi_{\rm eff}}{\partial z \partial R} \delta R
- \frac{\partial^2 \Phi_{\rm eff}}{\partial z^2}\delta z.
\label{vareq}
\end{eqnarray}

Consequently, the  corresponding Hamiltonian to the effective potential
given in Eq. (\ref{veff}) can be written as
\begin{equation}
H = \frac{1}{2} \left(\dot{R}^2 + \dot{z}^2 \right) + \Phi_{\rm eff}(R,z) = E,
\label{ham}
\end{equation}
where $\dot{R}$ and $\dot{z}$ are momenta per unit mass, conjugate to
$R$ and $z$ respectively, while $E$ is the numerical value of the
Hamiltonian, which is conserved.  Therefore, an orbit is restricted to
the area in the meridional plane satisfying $E \ geq \Phi_{\rm eff}$.

\sectionb{3}{COMPUTATIONAL METHODS}

\begin{figure}[!tH]
\centerline{\includegraphics[width=0.7\hsize]{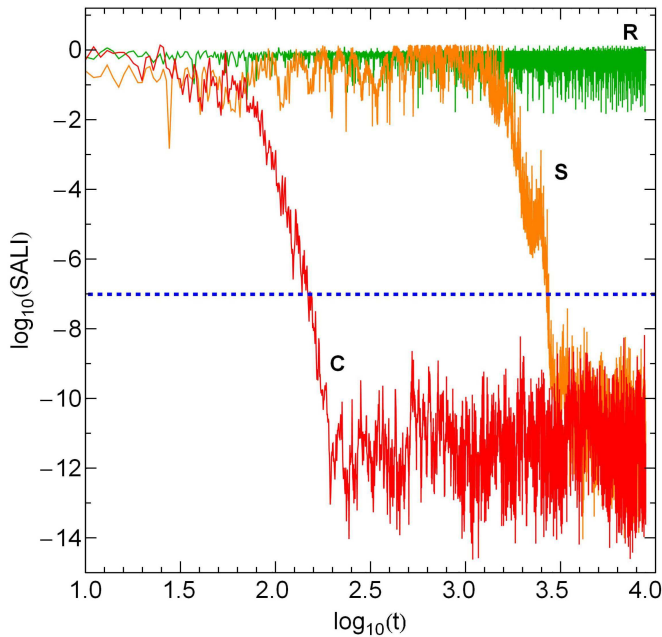}}
\caption{Time-evolution of the SALI of a regular orbit (green color
-- R), a sticky orbit (orange color -- S) and a chaotic orbit (red color
-- C) in our model for a time period of $10^4$ time units.  The
horizontal, blue dashed line corresponds to the threshold value
$10^{-7}$ which separates regular from chaotic motion.  The chaotic
orbit needs only about 150 time units in order to cross the threshold,
while the sticky orbit requires about 3000 time units so as to reveal
its true chaotic nature.}
\label{salis}
\end{figure}

In our numerical exploration, we seek to determine whether an orbit is
regular or chaotic.  Several indicators of chaos are available in the
literature; we chose the SALI indicator introduced in \citet{S01}.  The
time-evolution of SALI strongly depends on the nature of the computed
orbit since when the orbit is regular the SALI exhibits small
fluctuations around nonzero values, while on the other hand, in the case
of chaotic orbits the SALI after a small transient period it tends
exponentially to zero approaching the limit of the accuracy of the
computer $(10^{-16})$.  Therefore, the particular time-evolution of the
SALI allows us to distinguish fast and safely between regular and
chaotic motion.  The time-evolution of a regular (R) and a chaotic (C)
orbit for a time period of $10^4$ time units is presented in Fig.~\ref{salis}
 We observe, that both regular and chaotic orbits exhibit
the expected behavior.  Nevertheless, we have to define a specific
numerical threshold value for determining the transition from regularity
to chaos.  After conducting extensive numerical experiments, integrating
many sets of orbits, we conclude that a safe threshold value for the
SALI taking into account the total integration time of $10^4$ time units
is the value $10^{-7}$.  The horizontal, blue and dashed line in Fig.~\ref{salis}
corresponds to that threshold value which separates regular
from chaotic motion.  In order to decide whether an orbit is regular or
chaotic, one may use the usual method according to which we check after
a certain and predefined time interval of numerical integration, if the
value of SALI has become less than the established threshold value.
Therefore, if SALI $\leq 10^{-7}$ the orbit is chaotic, while if SALI $
> 10^{-7}$ the orbit is regular.

However, depending on the particular location of each orbit, this
threshold value can be reached more or less quickly, as there are
phenomena that can hold off the final classification of an orbit (i.e.,
there are special orbits called ``sticky" orbits, which behave
regularly for long time periods before they finally drift away from the
regular regions and start to wander in the chaotic domain, revealing
their true chaotic nature fully.  A characteristic example of a sticky
orbit (S) in our galactic model can be seen in Fig.~\ref{salis}, where
we observe that the chaotic character of the particular sticky orbit is
revealed only after a vast integration time of about 3000 time units.


To examine the orbital properties (chaoticity or regularity) of the
dynamical system, we need to establish some samples of initial
conditions of orbits.  The best approach, undoubtedly, would have been
to extract these samples of orbits from the distribution function of the
model.  Unfortunately, this is not available, so we followed another
course of action.  To determine the character of the orbits in our
model, we chose, for each value of the scale length $c_h$ of the dark
matter halo, a dense grid of initial conditions regularly distributed in
the area allowed by the value of the orbital energy $E$.  Our
investigation takes place in the phase space $(R,\dot{R})$ for a better
understanding of the orbital structure of the system.  The step
separation of the initial conditions along the $R$ and $\dot{R}$ axes
(in other words the density of the grids) was controlled in such a way
that always there are at least 50\,000 orbits to be integrated.  The
grids of initial conditions of orbits whose properties will be examined
are defined as follows:  we consider orbits with the initial conditions
$(R_0, \dot{R_0})$ with $z_0 = 0$, while the initial value of
$\dot{z_0}$ is obtained from the energy integral (\ref{ham}).  For each
initial condition, we numerically integrated the equations of motion
(\ref{eqmot}) as well as the variational equations (\ref{vareq}) with a
double precision Bulirsch-Stoer algorithm (e.g., \citealt{PTVF92}) with
a small time step of the order of $10^{-2}$, which is sufficient enough
for the desired accuracy of our computations (i.e., our results
practically do not change by halving the time step).  In all cases, the
energy integral (Eq.  \ref{ham}) was conserved better than one part in
$10^{-10}$, although for most orbits it was better than one part in
$10^{-11}$.

Each orbit was numerically integrated for a time interval of $10^4$ time
units ($10^{12}$ yr), which corresponds to a time span of the order of
hundreds of orbital periods.  The particular choice of the total
integration time is an element of great importance, especially in the
case of the sticky orbits.  A sticky orbit could be easily misclassified
as regular by any chaos indicator\footnote{~Generally, dynamical methods
are broadly split into two types:  (i) those based on the evolution of
sets of deviation vectors to characterize an orbit and (ii) those based
on the frequencies of the orbits that extract information about the
nature of motion only through the basic orbital elements without the use
of deviation vectors.}, if the total integration interval is too small,
so that the orbit does not have enough time to reveal its true chaotic
character.  Thus, all the initial conditions of the orbits of a given
grid were integrated, as we already said, for $10^4$ time units, thus
avoiding sticky orbits with a stickiness at least of the order of
$10^2$ Hubble time.  All the sticky orbits that do not show any signs of
chaoticity for $10^4$ time units are counted as regular orbits since
such vast sticky periods are completely out of the scope of our
research.

A first step toward the understanding of the overall behavior of our
system is knowing whether the orbits in the galactic model are regular
or chaotic.  Also of particular interest is the distribution of regular
orbits into different families.  Therefore, once the orbits have been
characterized as regular or chaotic, we then further classify the
regular orbits into different families by using a frequency analysis
method (\citealt{CA98,MCW05}).  Initially, \citet{BS82,BS84} proposed a
technique, dubbed spectral dynamics, for this particular purpose.
Later on, this method has been extended and improved by \citet{CA98} and
\citet{SN96}.  In a recent work, \citet{ZC13} the algorithm was refined
even further so it can be used to classify orbits in the meridional
plane.  In general terms, this method computes the Fourier transform of
the coordinates of an orbit, identifies its peaks, extracts the
corresponding frequencies, and searches for the fundamental frequencies
and their possible resonances.  Thus, we can easily identify
various families of regular orbits and also recognize the secondary
resonances that bifurcate from them.  This technique has been applied in
several previous papers (e.g., \citealt{ZC13,CZ13,ZCar13,ZCar14}) in the
field of orbit classification (not only regular versus chaotic, but
also separating regular orbits into different regular families) in
different galactic gravitational potentials.

Before closing this section, we would like to make a clarification about
the nomenclature of orbits.  All the orbits of an axisymmetric potential
are in fact three-dimensional (3D) loop orbits, i.e., orbits that always
rotate around the axis of symmetry in the same direction.  However, in
dealing with the meridional plane, the rotational motion is lost, so the
path that the orbit follows onto this plane can take any shape,
depending on the nature of the orbit.  Following the same approach of
the previous papers of this series, we characterize an orbit according
to its behavior in the meridional plane.  If, for example, an orbit is a
rosette lying in the equatorial plane of the axisymmetric potential, it
will be a linear orbit in the meridional plane, a tube orbit  will be
a 2:1 orbit, etc.  We should emphasize that we use the term ``box orbit"
for an orbit that conserves circulation, but this refers \textbf{only}
to the circulation provided by the meridional plane itself.  Because of
the their boxlike shape in the meridional plane, such orbits were
originally called ``boxes" (e.g., \citealt{O62}), even though their
three-dimensional shapes are more similar to doughnuts (see the review
of \citealt{M99}).  Nevertheless, we kept this formalism to maintain
continuity with all the previous papers of this series.

\sectionb{4}{ORBIT CLASSIFICATION}


\begin{figure}[!t]
\centerline{\includegraphics[height=170mm]{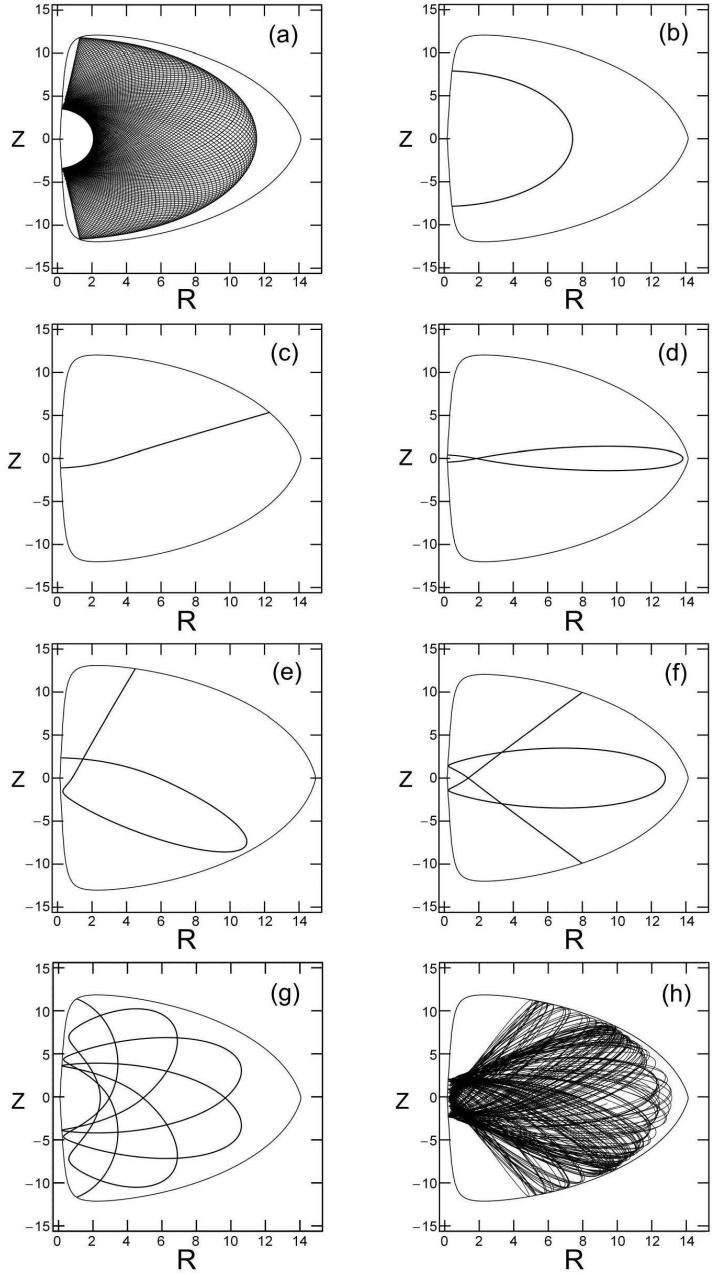}}
\caption{Orbit collection of the basic types of orbits in our galaxy
model:  (a) box orbit; (b) 2:1 banana-type orbit; (c) 1:1 linear orbit;
(d) 2:3 fish-type orbit; (e) 3:2 boxlet orbit; (f) 4:3 boxlet orbit; (g)
10:7 boxlet orbit, one of our ``orbits with higher resonance"; (h)
chaotic orbit. $R$ and $z$ are the cylindrical coordinates.}
\label{orbs}
\end{figure}

\begin{table}
{\small
\begin{center}
 \caption{Types and initial conditions of the
orbits shown in Fig.~\ref{orbs} (a-h).
   In all cases, $z_0 = 0$, $\dot{z_0}$ is found from
  the energy integral, Eq. (\ref{ham}), while $T_{\rm per}$ is the
period of the resonant parent periodic orbits.}
   \label{table}
   \begin{tabular}{@{}|c|c|r|r|r|}
      \hline
Figure & Type & $R_0$~~~~ & $\dot{R_0}$~~~~ & $T_{\rm per}$~~~~  \\
      \hline
      Fig.~3a &  box         &  2.10000000 &  0.00000000 &  --~~~~~~ \\
      Fig.~3b &  2:1 banana  &  7.39378964 &  0.00000000 &  1.90395532 \\
      Fig.~3c &  1:1 linear  &  3.39916301 & 42.85969675 &  1.28112540 \\
      Fig.~3d &  2:3 boxlet  & 13.75776855 &  0.00000000 &  2.55010488 \\
      Fig.~3e &  3:2 boxlet  &  0.85335785 & 20.81656490 &  3.94317967 \\
      Fig.~3f &  4:3 boxlet  & 12.74460663 &  0.00000000 &  5.12001897 \\
      Fig.~3g & 10:7 boxlet  &  2.48051157 &  0.00000000 & 12.99471468 \\
      Fig.~3h &  chaotic     &  0.25000000 &  0.00000000 &  --~~~~~~ \\
      \hline
   \end{tabular}
\end{center}
}
\end{table}

In this section, we will numerically integrate several sets of orbits,
in an attempt to distinguish the regular or chaotic nature of motion of
stars.  We use the sets of initial conditions described in the previous
section to construct the respective grids, always adopting values inside
the zero velocity curve (ZVC).  In all cases, the energy was set equal
to $-1000$, the angular momentum of the orbits is $L_{\rm z} = 10$,
while the scale length of the dark halo is treated as a parameter.
Here, we have to point out that the energy level controls the size of
the grid and particularly the $R_{\rm max}$ which is the maximum
possible value of the $R$ coordinate.  We chose that energy level $(E =
-1000)$ which yields 10 kpc $< R_{\rm max} < 17$ kpc.  To study how
scale length of the dark halo $c_{\rm h}$ influences the level of chaos,
we let it vary while fixing all other parameters of our galaxy model.
As already noted, we fix the values of all other parameters and
integrate orbits in the meridional plane for the set $c_{\rm h} =
\{5,7.5,10, ..., 30\}$.  Once the values of the parameters were chosen,
we computed a set of initial conditions as described in Section 3 and
integrated the corresponding orbits computing the SALI of the orbits and
then classifying regular orbits into different families.

Our numerical investigation reveals that in our galaxy model there are
eight basic types of orbits:  (i) chaotic orbits; (ii) box orbits; (iii)
1:1 linear orbits; (iv) 2:1 banana-type orbits; (v) 2:3 fish-type
orbits; (vi) 3:2 resonant orbits; (vii) 4:3 resonant orbits and (viii)
orbits with other resonances (i.e., all resonant orbits not included in
the former categories).  It turns out that for these last orbits the
corresponding percentage is less than 1\% in all cases, and therefore
their contribution to the overall orbital structure of the galaxy is
insignificant.  A $n:m$ resonant orbit would be represented by $m$
distinct islands of invariant curves in the phase plane $(R,\dot{R})$
and $n$ distinct islands of invariant curves in the $(z,\dot{z})$
surface of section.  In Fig.~\ref{orbs} (a-h) we present examples of
each of the basic types of regular orbits, plus an example of a chaotic
one.  In all cases, we set $c_{\rm h} = 15$ (except for the 3:2 resonant
orbit, where $c_{\rm h} = 12.5$).  The orbits shown in Figs.
\ref{orbs}a and \ref{orbs}h were computed until $t = 200$ time units,
while all the parent periodic orbits were computed until one period has
completed.  The black thick curve circumscribing each orbit is the
limiting curve in the meridional plane $(R,z)$ defined as $\Phi_{\rm
eff}(R,z) = E$.  Table \ref{table} shows the types and the initial
conditions for each of the depicted orbits; for the resonant cases, the
initial conditions and the period $T_{\rm per}$ correspond to the
parent\footnote{~For every orbital family there is a parent (or mother)
periodic orbit, i.e. an orbit that describes a closed figure.
Perturbing the initial conditions which define the exact position of a
periodic orbit we generate quasi-periodic orbits that belong to the
same orbital family and librate around their closed parent periodic
orbit.} periodic orbits.

At this point, we should note that the 1:1 resonance is usually the
hallmark of the loop orbits and both coordinates oscillate with the same
frequency in their main motion.  Their mother orbit is a closed loop
orbit.  Moreover, when the oscillations are in phase, the 1:1 orbit
degenerates into a linear orbit (the same as in Lissajous figures made
with two oscillators).  In our meridional plane, however, 1:1 orbits do
not have the shape of a loop.  In fact, their mother orbit is linear
(e.g., Fig.~\ref{orbs}c) , and thus they do not have a hollow (in the
meridional plane), but fill a region around the linear mother, always
oscillating along the $R$ and $z$ directions with the same frequency.
We designate these orbits ``1:1 linear open orbits" to differentiate
them from true meridional plane loop orbits, which have a hollow and
also always rotate in the same direction.


\begin{figure}[!t]
\centerline{\resizebox{\hsize}{!}{\includegraphics{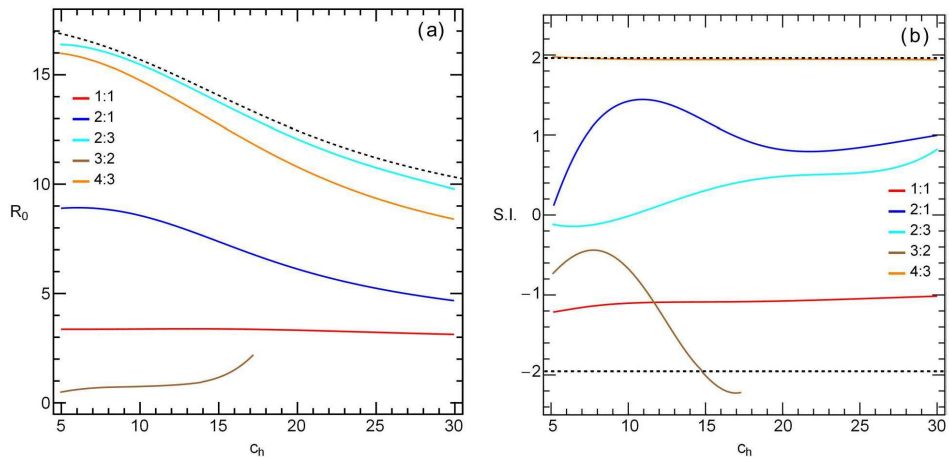}}}
\caption{Panel (a), left:  the $(R_0,c_{\rm h})$ characteristic curves of
the orbital families; panel (b), right:  evolution of the stability index S.I.
of the families of periodic orbits shown in panel (a).  The black
horizontal dashed lines at --2 and +2 delimit the range of S.I. for
which the periodic orbits are stable.}
\label{otd}
\end{figure}

In Fig.~\ref{otd}a we present a very informative diagram the so-called
``characteristic" orbital diagram \citet{CM77}.  It shows the evolution
of the $R$ coordinate of the initial conditions of the parent periodic
orbits of each orbital family as a function of the variable scale
length of the dark halo $c_{\rm h}$.  Here we should emphasize, that for
orbits starting perpendicular to the $R$-axis, we need only the initial
condition of $R_0$ in order to locate them on the characteristic
diagram.  On the other hand, for orbits not starting perpendicular to
the $R$-axis (i.e, the 1:1 and 3:2 resonant orbits) initial conditions
as position-velocity pairs $(R,\dot{R})$ are required and, therefore,
the characteristic diagram is three-dimensional providing  full
information regarding the interrelations of the initial conditions in a
tree of families of periodic orbits.  The outermost black dashed line
denotes the permissible area of motion as it is defined by the
particular value of the energy; the region above this line is
forbidden.  Furthermore, the diagram shown in Fig.~\ref{otd}b is called
the ``stability diagram" (\citealt{CB85,CM85}), and it illustrates the
stability of all the families of periodic orbits in our dynamical system
when the numerical value of $c_{\rm h}$ varies, while all other
parameters remain constant.  A periodic orbit is stable if only the
stability index (S.I.)  (\citealt{MH92,Z13}) is between --2 and +2.
This diagram help us to monitor the evolution of S.I. of the resonant
periodic orbits as well as the transitions from stability to instability
and vice versa.  Our computations suggest that almost all resonant
families are stable throughout the interval $c_{\rm h} \in [5, 30]$.  We
observe that the curve of the 4:3 is located very close to the upper
stability limit, evolving asymptotically, but in never crosses the
critical value (+2), at least inside the studied interval of values of
$c_{\rm h}$.  Contrary, the 3:2 family is stable only for $c_{\rm
h} < 14.71$, while for larger values of the dark halo scale length the
resonant periodic orbits become unstable and eventually the 3:2 family
terminates when $c_{\rm h} = 17.1923$.


\begin{figure}[!t]
\centerline{\includegraphics[height=175mm]{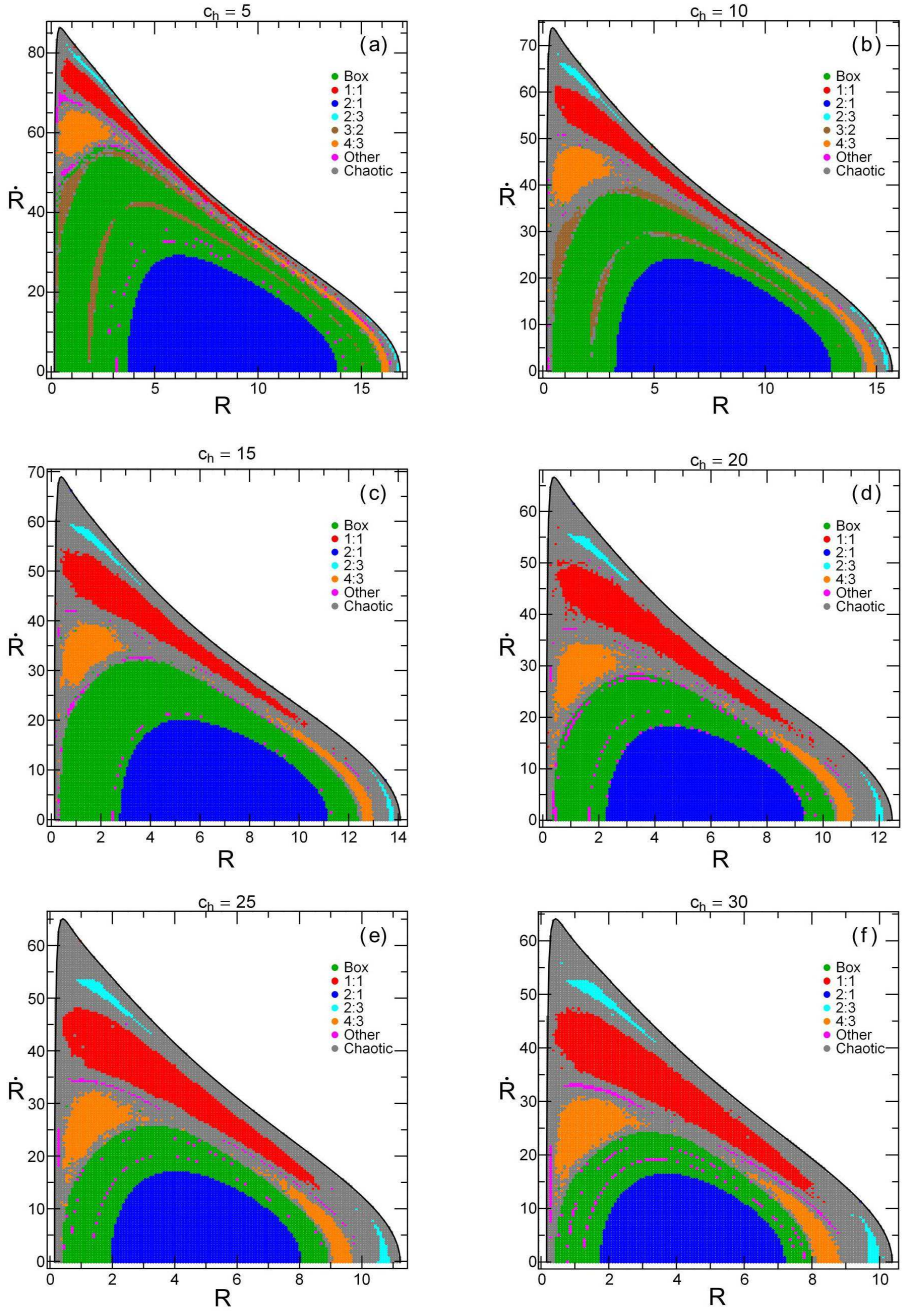}}
 \caption{Orbital structure of the phase plane $(R,\dot{R})$  of our
 galaxy model for different values of the scale length of the dark halo
$c_{\rm h}$.}
\label{grids}
\end{figure}

In Figs.~\ref{grids} (a--f) we present six grids of initial conditions
$(R_0,\dot{R_0})$ of orbits that we have classified for different values
of the scale length of the dark matter halo $c_{\rm h}$.  These
color-coded grids of initial conditions are equivalent to the classical
Poincar\'{e} surfaces of section (PSS) and allow us to determine what
types of orbits occupy specific areas in the phase plane $(R,\dot{R})$.
The outermost black thick curve is the limiting curve which is defined
as
\begin{equation}
\frac{1}{2} \dot{R}^2 + \Phi_{\rm eff}(R,0) = E.
\label{zvc}
\end{equation}

In Fig.~\ref{grids}a we see that when the spherical dark matter halo is
very concentrated $(c_{\rm h} = 5)$, the vast majority of the phase
plane is covered by initial conditions corresponding to regular orbits,
while chaos is confined only at the outer parts of the phase plane thus
displaying a thin chaotic layer.  In particular, we can distinguish the
seven main types of regular orbits presented earlier:  (i) 2:1
banana-type orbits located at the central region of the grid; (ii) box
orbits situated mainly outside of the 2:1 resonant orbits; (iii) 1:1
open linear orbits form the elongated island of initial conditions; (iv)
2:3 fish-type resonant orbits form the set of three\footnote{~It should
be pointed out that the color-coded grids of Fig.~\ref{grids} (a--f )
show only the $\dot{R} > 0$ part of the phase plane; the $\dot{R} < 0$
is symmetrical.  Therefore, in many resonances not all the corresponding
stability islands are present (e.g., for the 2:3 and 4:3 resonances only
two of the three islands are shown).} small islands at the outer parts
of the grid; (v) 3:2 resonant orbits producing several stability islands
inside the box region; (vi) 4:3 resonant orbits form the chain of three
islands; and (vii) other types of resonances producing extremely small
islands embedded both in the chaotic and box areas.  However, as the
value of scale length increases and the dark matter halo becomes less
and less concentrated, we observe that the amount of box and 2:1
resonant orbits reduces thus giving place to other resonant orbits
(i.e., the 1:1, 2:3 and 4:3 families) to grow, while at the same time
the chaotic region increases in size.  When $c_{\rm h} = 15$, Fig.~\ref{grids}c
shows that there is no indication whatsoever of the 3:2
family.  This is expected because for $14.71 < c_{\rm h} < 17.1923$ the
3:2 resonant orbits are unstable (see Fig.~\ref{otd}, a-b).  This means
that the periodic point of the 3:2 resonance is indeed present in
Fig.~\ref{grids}c,
although evidently deeply buried in the box area.  For
relatively large values of the scale length $(c_{\rm} \geq 20)$ one may
see in Figs.~\ref{grids} (d--f) that there is a substantial chaotic
region, while higher resonant orbits (i.e., the 4:5, 6:5, 7:5 and 10:7)
produce thin filaments of initial conditions inside the box an chaotic
regions.  It should also be mentioned that the permissible area on the
phase plane (both the $R_{\rm max}$ and the radial velocity $\dot{R}$ of
the stars near the center of the galaxy are reduced) is reduced with
increasing scale length of the dark matter halo.


\begin{figure}[!tH]
\centerline{\includegraphics[width=0.60\hsize]{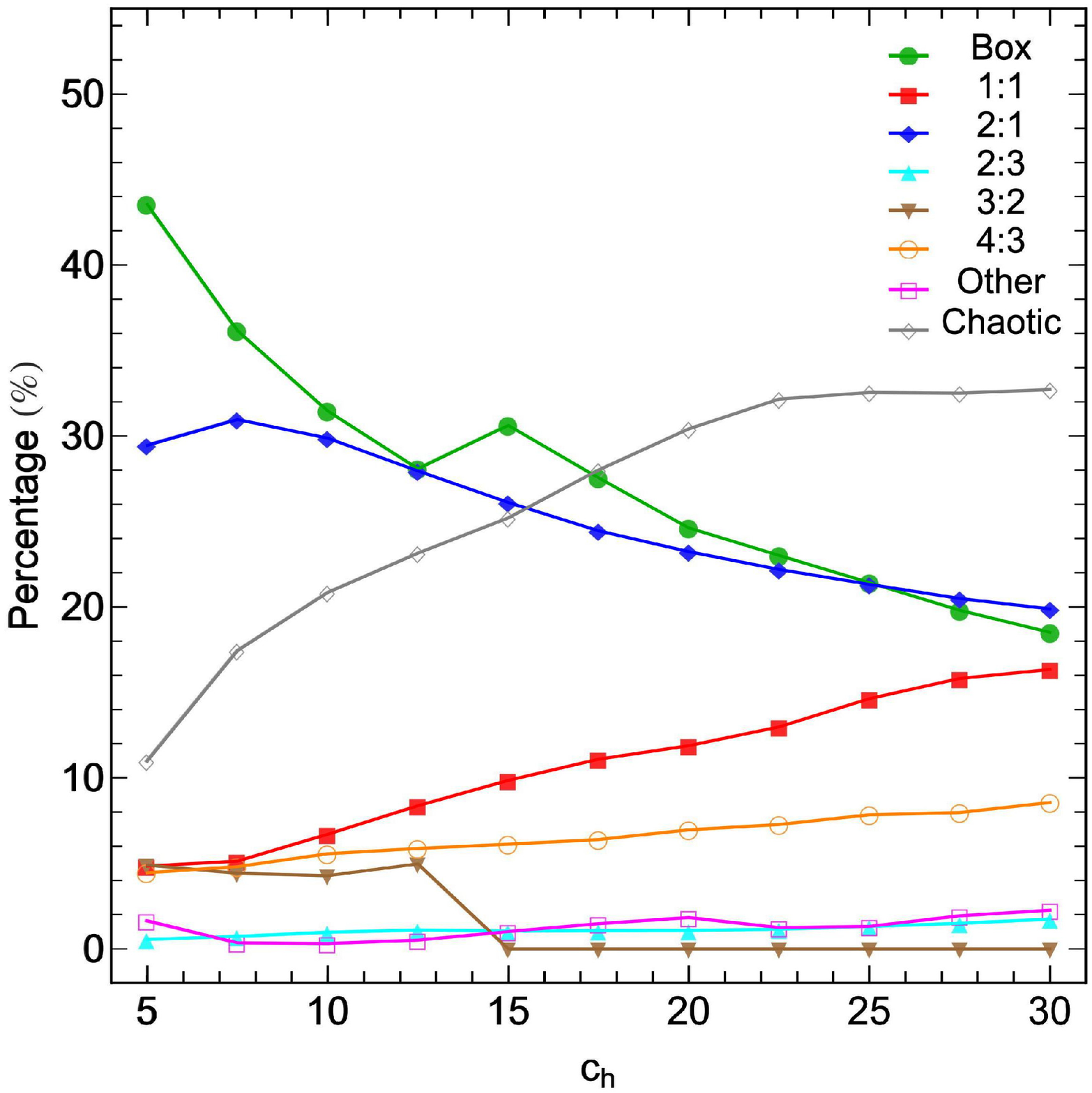}}
 \caption{Evolution of the percentages of different types of
 orbits in the phase plane $(R,\dot{R})$ of our galaxy model, when
varying the value of the scale length of the dark halo $c_{\rm h}$.}
\label{percs}
\vspace{5mm}
\centerline{\includegraphics[width=0.62\hsize]{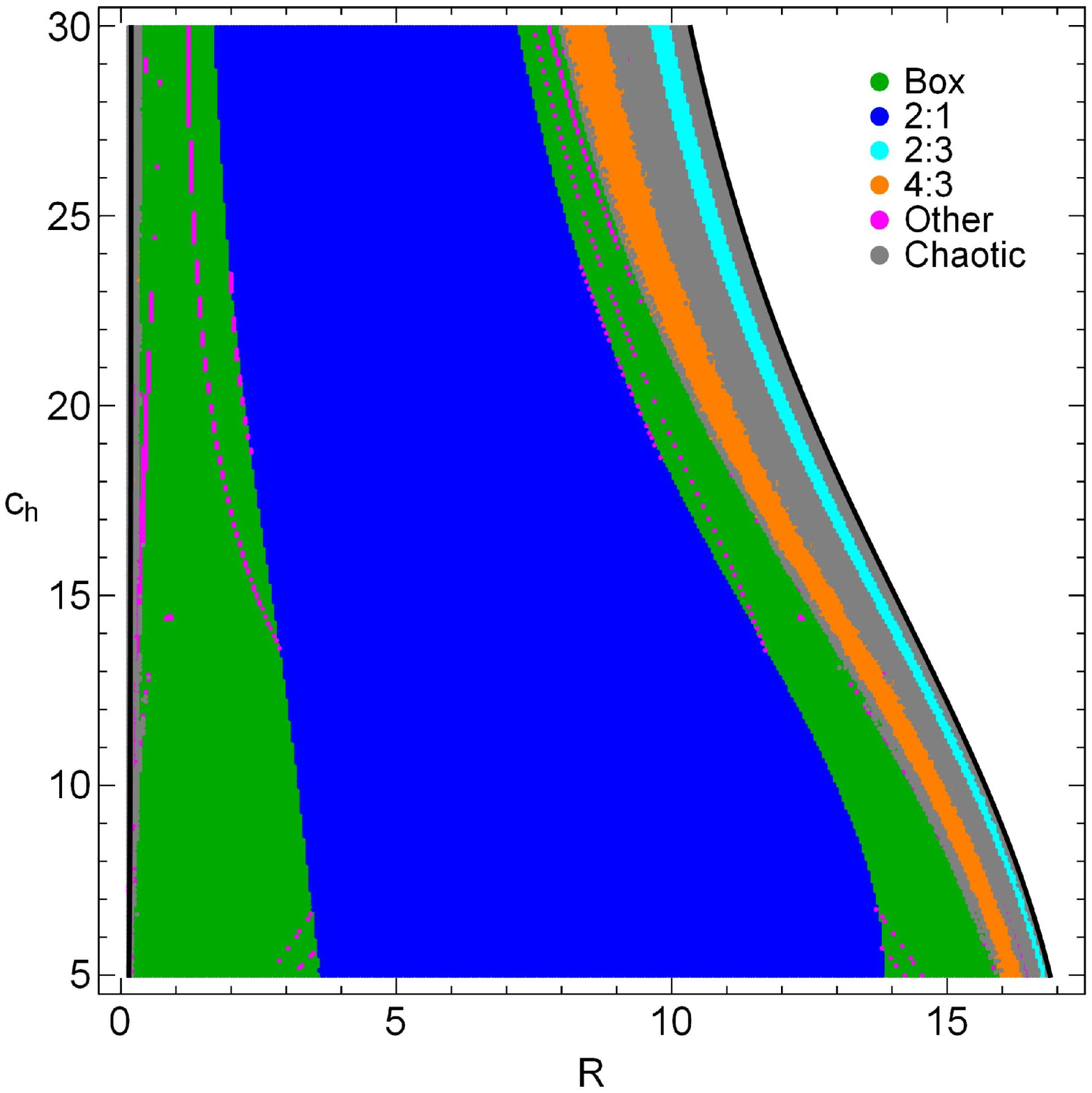}}
\caption{Orbital structure of the $(R,c_{\rm h})$ plane.  This
diagram gives a detailed analysis of the evolution of orbits starting
perpendicularly to the $R$-axis when the value of the halo scale length
varies in the interval $c_{\rm h} \in [5, 30]$.}
\label{Rch}
\end{figure}

The following Fig.~\ref{percs} shows the evolution of the percentages
of the chaotic and all types of regular orbits as a function of the
scale length of the dark matter halo, $c_{\rm h}$.  One may observe
that, as we proceed to higher values of the scale length, the percentage
of chaotic orbits increases, while at the same time, the rate of box and
2:1 resonant orbits reduces steadily.  In the case of very low scale
length $(c_{\rm h} = 5)$ almost the entire phase plane is covered only
by initial conditions of regular orbits; about 45\% of the phase plane
corresponds to initial conditions of box orbits and only about 10\% of
it to chaotic orbits.  However, when $c_{\rm h} > 17.5$, chaotic orbits
is the most populated family dominating the phase plane.  It is also
seen that, when $c_{\rm h} > 22.5$, the percentage of chaotic orbits
seem to saturate around 32\%.  At the highest studied value of the scale
length $(c_{\rm h} = 30)$ or in other words when the dark halo is very
loose, the rates of box, 2:1 and 1:1 resonant orbits tend to a common
value (around 20\%), thus sharing three fifths of the phase plane.
Furthermore, our numerical analysis suggests that the percentages of the
1:1, 2:3 and 4:3 families grow with increasing scale length, although
following different rates.  In fact, the amount of 1:1 resonant orbits
increases almost linearly, while the 2:3 and 4:3 families exhibit a less
sharp increase, so we may argue that these types of orbits are
practically immune to the change of $c_{\rm h}$.  Moreover, for $c_{\rm
h} < 12.5$, the 3:2 family holds a constant rate around 5\%, while for
larger values of $c_{\rm h}$ it vanishes.  In addition, we may say that
in general terms, all other resonant families possess throughout very
low percentages (always less than 5\%) so, varying the value of $c_{\rm
h}$ only shuffles the orbital content among them.  Thus, taking into
account all the above-mentioned analysis, we may conclude that in the
phase plane $(R,\dot{R})$ the types of orbits, that are mostly
influenced by the scale length of the spherical dark matter halo, are
the box, 1:1, 2:1 and the chaotic orbits.

The grids in the phase plane $(R,\dot{R})$ can provide information on
the phase space mixing for only a fixed value of the halo scale length
$c_{\rm h}$.  However, H\'{e}non back in the 60s considered a plane
which provides information about regions of regularity and regions of
chaos using the section $z = \dot{R} = 0$, $\dot{z} > 0$, i.e., the test
particles (stars) are launched on the $R$-axis, parallel to the $z$-axis
and in the positive $z$-direction.  Thus, in contrast to the previously
discussed grids, only orbits with pericenters on the $R$-axis are
included and therefore the value of $c_{\rm h}$ is now used as an
ordinate.  Fig.~\ref{Rch} shows the orbital structure of the $(R,c_{\rm
h})$-plane when $c_{\rm h} \in [5, 30]$.  In order to be able to monitor
with sufficient accuracy and details the evolution of the families of
orbits, we defined a dense grid of $10^5$ initial conditions in the
$(R,c_{\rm h})$-plane.  It is evident, that the vast majority of the
grid is covered either by box or 2:1 resonant orbits, while initial
conditions of chaotic orbits are mainly confined to right outer part of
the $(R,c_{\rm h})$-plane.  Furthermore, the 2:3 and 4:3 resonances
produce thin stability layers.  It is also seen, that several families
of higher resonant orbits are present, corresponding to thin filaments
of initial conditions living inside the box region.  It should be
emphasized, that the maximum value of the $R$ coordinate $(R_{\rm max})$
is reduced with increasing $c_{\rm h}$.  We must also stress out that
the $(R,c_{\rm h})$-plane contains only such orbits starting
perpendicularly to the $R$-axis, while orbits whose initial conditions
are pairs of position-velocity (i.e., the 1:1 and 3:2 resonant families)
are obviously not included.

\sectionb{5}{DISCUSSION AND CONCLUSIONS}

The aim of the present work was to investigate how influential is the
scale length of the dark matter halo on the level of chaos and on the
distribution of regular families among its orbits.  For this purpose, we
used an analytic, axially symmetric galactic gravitational model which
embraces the general features of a disk galaxy with a dense, massive,
central nucleus and an additional spherical dark matter halo component.
To simplify our study, we chose to work in the meridional plane $(R,z)$,
thus reducing three-dimensional to two-dimensional motion.  We kept the
values of all other parameters constant, because our main objective
was to determine the influence of the scale length on the percentages of
the orbits.  Our thorough and detailed numerical analysis suggests that
the level of chaos as well as the distribution in regular families is
indeed very dependent on the halo scale length.

Undoubtedly, a disk galaxy with a central massive nucleus and a dark
matter halo is a very complex entity and therefore, we need to assume
some necessary simplifications and assumptions in order to be able to
study mathematically the orbital behavior of such a complicated stellar
system.  For this purpose, our total gravitational model is
intentionally simple and contrived, in order to give us the ability to
study all different aspects regarding the kinematics and dynamics of the
model.  Nevertheless, contrived models can surely provide an insight
into more realistic stellar systems, which unfortunately are very
difficult to be explored, if we take into account all the astrophysical
aspects (i.e., gas, spirals, mergers, etc.).  On the other hand,
self-consistent models are mainly used when conducting $N$-body
simulations.  However, this application is entirely out of the scope of
the present paper.  Once more we have to point out that the simplicity
of our model is necessary, otherwise it would be extremely difficult, or
even impossible, to apply the extensive and detailed numerical
calculations presented in this study.  Similar gravitational models with
the same limitations and assumptions were used successfully several
times in the past in order to investigate the orbital structure in much
more complicated galactic systems.

Our numerical investigation takes place in the phase space $(R,\dot{R})$
for a better understanding the orbital structure of the system.  Since a
distribution function of our galaxy model was not available for
extracting the different samples of orbits, we had to follow an
alternative path.  Thus, for determining the regular or chaotic
character of orbits in our models, we defined dense grids of initial
conditions regularly distributed in the area allowed by the value of the
total orbital energy $E$, in the phase space.  To show how the halo
scale length influences the orbital structure of the system, for each
case we presented color-coded grids of initial conditions, which allow
us to visualize what types of orbits occupy specific areas in the phase
space.  Each orbit was integrated numerically for a time period of
$10^4$ time units ($10^{12}$ yr), which corresponds to a time span of
the order of hundreds of orbital periods.  The particular, choice of the
total integration time was made in order to eliminate sticky orbits
(classifying them correctly as chaotic orbits) with a stickiness at
least of 100 Hubble times.  Then, we made a step further in an attempt
to distribute all regular orbits into different families.  Therefore,
once an orbit has been characterized as regular applying the SALI
method, we then further classified it using a frequency analysis method.
This method calculates the Fourier transform of the coordinates and
velocities of an orbit, identifies its peaks, extracts the corresponding
frequencies and then searches for the fundamental frequencies and their
possible resonances.

The main outcomes of our research can be summarized as follows:

$\bullet$ Several types of regular orbits were found to exist
in our galactic gravitational model, while there are also extended
chaotic domains separating the areas of regularity.  In particular, a
large variety of resonant orbits (i.e., 1:1, 2:1, 2:3, 3:2, 4:3 and
higher resonant orbits) are present, thus making the orbital structure
more rich.  Here we must clarify that by the term ``higher resonant
orbits" we refer to resonant orbits with a rational quotient of
frequencies made from integers $>$\,5, which of course do not belong to
the main families.

$\bullet$ It was found that in the phase space $(R,\dot{R})$
the scale length of the dark matter halo influences mainly box, 1:1, 2:1
and chaotic orbits.  Moreover, the majority of stars move in regular
orbits and in general terms, box and 2:1 resonant orbits are the most
populated family of regular orbits throughout the range of values of
$c_{\rm h}$.  All resonant families are stable apart from the 3:2 family
which consists of a mixture of stable and unstable resonant orbits.

$\bullet$ The largest amount of chaos, about 32\%, was
measured for high values of the scale length $(c_{\rm h} > 22.5)$
corresponding to very loose dark matter haloes, while as $c_{\rm h}$
decreases and the dark halo becomes more and more concentrated, the
amount of chaotic orbits is reduced rapidly, and for the low enough
values of the scale length, $(c_{\rm h} < 7)$, about 90\% of the tested
orbits were found to be regular.

The influence of the scale length of the dark halo on the character of
orbits has been an active field of investigation over the years.  In
\citet{CZ11} the dark mater halo was also modeled by a mass Plummer type
potential and it was observed that the amount of chaos in triaxial
galaxies is higher when they are surrounded by less concentrated dark
matter haloes.  This behavior is in absolute agreement with the current
results.  A logarithmic potential has been used in \citet{Z14} to
describe the properties of a biaxial dark matter halo.  There it was
found that in galaxy models with prolate or oblate dark matter haloes,
the scale length of the halo affects mostly box, 2:1 banana-type,
resonant and chaotic orbits, while a similar relation between the
amount of chaos an the scale length was obtained.  Therefore we may
conclude that the scale length, which is the parameter controlling how
concentrated is the dark matter halo, has roughly the same impact on the
character of orbits in galaxies regardless the particular type of the
potential used for modeling the dark matter halo.

\end{document}